\documentclass[12pt]{article}

\begin{document}
\begin{center}
{\bf Pair Production of Arbitrary Spin
Particles by Electromagnetic Fields}\\
\vspace{5mm}
 S.I. Kruglov \\
\vspace{5mm}
\textit{International Educational Centre, \\
Toronto, Ontario, Canada L4J 1E3}
\end{center}

\begin{abstract}
The exact solutions of the wave equation for arbitrary spin
particles in the field of the soliton-like electric impulse were
obtained. The differential probability of pair production of
particles by electromagnetic fields has been evaluated on the
basis of the exact solutions. As a particular case, the particle
pair producing in the constant and uniform electric field were
studied.
\end{abstract}

\section{Introduction}

One of the most interesting nonlinear phenomena in quantum field
theory is the $e^+ e^- $-pair production by a static and uniform
electric field \cite{Schwinger}. It is impossible to describe this
effect in the framework of the perturbation theory which is
nonperturbative effect occurring in strong external fields. With
the aid of the semiclassical imaginary-time method, which is valid
only for weak electromagnetic fields, the pair production
probability of arbitrary spin particles was studied in
\cite{Marinov72} for the case of constant and uniform fields. We
calculated the probability of pair production of arbitrary spin
particles with the electric dipole moments (EDM) and anomalous
magnetic moments (AMM) by constant and uniform electromagnetic
fields of arbitrary strength \cite{Kruglov2001}. Spontaneous $e^+
e^- $-pair creation requires the strong constant and uniform
electric field of the value (critical field) $E_c\simeq 1.3\times
10^{18}~V/m$ that is difficult to achieve experimentally. In the
proposed X-ray free electron laser facilities at SLAC and DESY
\cite{SLAC}, \cite{Ringwald} the strength of the field can reach
$E\simeq 0.1E_c$. It is well known, however, that one laser beam,
which is described by a plane electromagnetic wave, can not
produce pairs of particles \cite{Schwinger}. With the help of two
coherent laser beams, it possible to form a standing wave which
produces pairs. Such a possibility of $e^+e^-$-pair creation by
X-ray free electron laser is discussed in \cite{Ringwald},
\cite{Alkofer}, \cite{Popov}. The radiation field of two crossing
laser beams can be approximated by spatial uniform and time
varying electric field \cite{Alkofer}. If the laser pulse duration
decreases the probability of electron-positron creation increases
sharply \cite{Popov}. Therefore, it is interesting to investigate
the general case of the pair production of arbitrary spin
particles by the electric impulse of the electromagnetic field.

The paper is organized as follows. In Section 2, the formulation
of arbitrary spin particles on the basis of the Lorentz
representation
$(s,0)\oplus(s-1/2,1/2)\oplus(0,s)\oplus(1/2,s-1/2)$ ($\oplus$
means the direct sum) in the form of the first order relativistic
wave equation is briefly considered. The spin operator, which
commutes with the equation matrix, is defined. The exact solutions
of the wave equation and the differential probability of pair
production of arbitrary spin particles in the field of the
soliton-like electric impulse are investigated in Section 3. In
Section 4, the case of the particle pair producing in the constant
and uniform electric field is studied. We discuss the results
obtained in Section 5.

We use units in which $\hbar =c=1$, and four-vectors are defined
as $V_\mu=(\textbf{V},V_4)$, $\textbf{V}=(V_m$) ($m=1,2,3$),
$V_4=iV_0$. The scalar product is  $\textbf{VP}=V_m P_m$.  The
Greek and Latin letters run numbers $1,2,3,4$, and $1,2,3$,
respectively. A summation is implied over the repeated indices.

\section{Arbitrary spin particles}

We consider the first order relativistic wave equation (RWE) for
arbitrary spin particles suggested in \cite{Hurley}. In an
external electromagnetic field this RWE takes the form
\begin{equation}
\left(\beta _\mu  \mathcal{D}_\mu + m\right) \phi_\epsilon (x)=0 ,
\label{1}
\end{equation}
where $\mathcal{D}_\mu =\partial_\mu -ieA_\mu$,
$\partial_\mu=(\partial/\partial_m,\partial/i\partial t)$, the $t$
is the time, $A_\mu=(\textbf{A},iA_0)$ is four-vector-potential.
The matrices and the wave function of RWE (1) are given by (in our
notation)
\[
 \beta _m =i\epsilon g\left(
\begin{array}{ccc}
0 & S_m & -K_m^+  \\
-S_m & 0 & 0  \\
K_m & 0 & 0 \\
\end{array}
\right) ,~~~~\beta_4=\left(
\begin{array}{ccc}
0 & 1 & 0  \\
1 & 0 & 0  \\
0 & 0 & 0  \\
\end{array}
\right) ,
\]
\vspace{-8mm}
\begin{equation}  \label{2}
\end{equation}
\vspace{-8mm}
\[
\phi_\epsilon (x)=\left(\begin{array}{ccc}
\psi_\epsilon (x) \\
\chi_\epsilon (x) \\
\Omega_\epsilon (x) \\
\end{array}\right) ,
\]
where $g$ is a gyromagnetic ratio, $g=1/s$ ($s$ is a spin of
particles), $K_m^+$ is Hermitian-conjugate matrix, and the
parameter $\epsilon=1$ corresponds to the
$(s,0)\oplus(s-1/2,1/2)$, and $\epsilon=-1$ to the
$(0,s)\oplus(1/2,s-1/2)$ representations of the Lorentz group.
Thus, the magnetic moment of arbitrary spin particles is $\mu
=\mu_B=e/(2m)$, $\mu_B$ is the Bohr magneton. The square
$(2s+1)\times (2s+1)$-dimensional spin matrices $S_m$ and
rectangular $(2s-1)\times (2s+1)$-dimensional matrices $K_m$ obey
the relationships \cite{Hurley}
\[
\left[ S_i,S_j\right] =i\epsilon _{ijk}S_k,\hspace{0.2in}\left(
S_1\right) ^2+\left( S_2\right) ^2+\left( S_3\right) ^2=s(s+1) ,
\]
\vspace{-8mm}
\begin{equation}  \label{3}
\end{equation}
\vspace{-8mm}
\[
S_iS_j+K_i^+ K_j=is\epsilon_{ijk}S_k +s^2\delta_{ij}
\hspace{0.2in}(i,j,k=1,2,3) ,
\]
where $\epsilon_{ijk}$ is antisymmetric tensor,
$\epsilon_{123}=1$. The functions $\psi_\epsilon (x)$,
$\chi_\epsilon (x)$ possess $2s+1$ components, and the
$\Omega_\epsilon (x)$ has $2s-1$ components, so that the wave
function $\phi_\epsilon(x)$ possesses $6s+1$ components. The
system of two equations (1) for $\epsilon =1$ (the wave function
$\phi_{+} (x)$) and $\epsilon =-1$ (the wave function $\phi_{-}
(x)$) is a parity invariant because at the $P$-transformation, the
representation of the Lorentz group $(s,0)\oplus(s-1/2,1/2)$ is
transformed into $(0,s)\oplus(1/2,s-1/2)$ representation. The
whole wave function $\phi(x)=\phi_{+} (x)\oplus \phi_{-} (x)$ has
$2(6s+1)$ components.

Here we do not discuss the case of arbitrary EDM and AMM of spin
particles that requires the consideration of high dimension RWE of
the first order (see \cite{Kruglov}).

Equation (1) is form-invariant under the Lorentz transformation
due to the relation \cite{Hurley}
\begin{equation}
\left[M_{\mu\nu},\beta_\sigma\right] =\delta_{\nu\sigma}\beta_\mu
- \delta_{\mu\sigma}\beta_\nu , \label{4}
\end{equation}
where $M_{\mu\nu}$ are generators of the Lorentz group, and the
commutator is defined as usual: $[A,B]=AB-BA$.

Now we construct the spin operator of a particle in external
electromagnetic field. In this paper we investigate particles in
electromagnetic fields that are spatially homogenies and dependent
only on time. Therefore, consider the electromagnetic
vector-potential of the form $A_\mu=A_\mu (t)$. The solution of
Eq. (1) can be represented as
\begin{equation}
\phi_\epsilon (x)=\Phi_\epsilon (t)\exp(i\textbf{p} \textbf{x})
,\label{5}
\end{equation}
where $\textbf{p}$ is the momentum vector and the function
$\Phi(t)$ obeys the equation
\begin{equation}
\left[i\beta _m\left(p_m-eA_m (t)\right)  -i\beta_4\left(
\partial_t +eA_4 (t)\right) + m\right] \Phi_\epsilon (t)=0 , \label{6}
\end{equation}
where $\partial_t=\partial/\partial t$.

The spin operator of particles corresponding to the time-dependent
vector-potential may be defined in the form
\begin{equation}
\hat{S}_ {\textbf{p}}=\frac{\textbf{J}\left( \textbf{
p}-e\textbf{A}(t)\right)}{|\textbf{p}-e\textbf{A}(t)|}
,~~~~J_m=\frac{1}{2}\epsilon_{mnk}M_{nk} = \left(
\begin{array}{ccc}
S_m & 0 & 0  \\
0 & S_m & 0  \\
0 & 0 & \Sigma_m \\
\end{array}\right) ,\label{7}
\end{equation}
where $\Sigma_m$ are generators of SU(2) group in the (2s-1) -
representation \cite{Hurley}. We consider the case when the
electric field, $\textbf{E}$, vanishes at $t\rightarrow \pm \infty
$. It is easy to verify with the help of Eq. (4) that the spin
operator $\hat{S}_ {\textbf{p}}$ (7) commutes with the matrix of
Eq. (6) for asymptotic states:
\[
[ \hat{S}_ {\textbf{p}},\Lambda]=0 ~~~~~~ (t\rightarrow \pm
\infty) ,
\]
\vspace{-8mm}
\begin{equation} \label{8}
\end{equation}
\vspace{-8mm}
\[
\Lambda\equiv i\beta _m\left(p_m-eA_m (t)\right) -i\beta_4\left(
\partial_t +eA_4 (t)\right) + m .
\]
Then the wave function $\Phi_\epsilon (t)$ (and $\phi_\epsilon
(x)$) is eigenfunction of the spin operator for asymptotic states:
\begin{equation}
\hat{S}_{\textbf{p}}\Phi_\epsilon (t)=s_z\Phi_\epsilon (t)
~~~~t\rightarrow \pm\infty ,\label{9}
\end{equation}
where $s_z$ is a spin projection, $s_z=-s,...s$.

The second order wave equation for the function $\psi _\varepsilon
(x)$ follows from Eq. (1), and is given by \cite{Hurley}:
\begin{equation}
\left[ \mathcal{D}_\mu ^2-m^2-\frac e2 gF_{\mu \nu }\Sigma _{\mu
\nu }^{(\epsilon )}\right] \psi _\epsilon (x)=0 , \label{10}
\end{equation}
where $F_{\mu \nu }=\partial _\mu A_\nu -\partial _{_\nu }A_\mu $
is the strength tensor of electromagnetic fiels. Eq. (10) may be
considered as the arbitrary - spin generalization of the
Feynman--Gell-Mann \cite{Feynman} equation (see also
\cite{Broun}). The generators $\Sigma_{\mu \nu }^{(\varepsilon )}$
of the representations $(s,0)$ for $\varepsilon =1$ and $(0,s)$
for $\varepsilon =-1$ are
\begin{equation}
\Sigma _{ij}^{(\epsilon )}=\epsilon _{ijk}S_k,\hspace{0.5in}\Sigma
_{0k}^{(\epsilon )}=-i\epsilon S_k .  \label{11}
\end{equation}
The components $\chi_\epsilon (x)$, $\Omega_\epsilon (x)$ of the
whole wave function $\phi_\epsilon$, (2), read
\begin{equation}
\chi_\epsilon (x)=\frac{1}{m}\left[ \left(i\partial_t -eA_0
\right)+i\epsilon g\left( \textbf{S}\textbf{D} \right)\right]\psi
_\epsilon (x) ,~~~~\Omega_\epsilon (x)=-\frac{\epsilon g}{m}\left(
\textbf{K}\textbf{D} \right)\psi _\epsilon (x) . \label{12}
\end{equation}
So, if one finds a solution $\psi _\varepsilon (x)$ of the second
order Eq. (10), the functions $\chi_\epsilon (x)$,
$\Omega_\epsilon (x)$ can be obtained from Eqs. (12).

In order to define a relativistically invariant bilinear form, we
introduce the $(12s+2)$-dimensional wave function \cite{Hurley}
\begin{equation}
\phi (x)=\left(
\begin{array}{c}
\phi _{+}(x) \\
\phi _{-}(x)
\end{array}
\right) .   \label{13}
\end{equation}
Then the Lorentz-invariant bilinear form is given by \cite{Hurley}
\begin{equation}
\overline{\phi }(x)\phi (x)=\phi ^{+}(x)\eta \phi (x) =\phi_+
^{+}(x)\phi_-(x)+\phi_- ^{+}(x)\phi_+(x),~~~~\eta=\left (
\begin{array}{cc}
0& 1 \\
1& 0
\end{array}
\right) , \label{14}
\end{equation}
where $\phi ^{+}(x)$ is the Hermitian-conjugate wave function.
From Eqs. (5),(7), (9), (12) one may find the relation for the
asymptotic state ($t\rightarrow \pm\infty$)
\[
\phi^+ _{+}(x)  \phi _{-}(x)=\psi^+_+(t)\biggl[1
-\frac{1}{m^2}\left(i \overleftarrow{\partial_t
}+eA_0(t)\right)\left(i \partial_t -eA_0(t)\right)
\]
\vspace{-8mm}
\begin{equation}  \label{15}
\end{equation}
\vspace{-8mm}
\[
-i\frac{gs_z}{m^2}\left(\overleftarrow{\partial_t
}|\textbf{p}-e\textbf{A}(t)|+|\textbf{p}-e\textbf{A}(t)|\partial_t\right)
-\frac{\left(\textbf{p}-e \textbf{A}(t) \right)^2}{m^2}
\biggr]\psi_-(t) ,
\]
where the derivative $ \overleftarrow{\partial_t}$ acts on the
left standing function. The second term in Eq. (14), $\phi_-
^{+}(x)\phi_+(x)=\left(\phi^+ _{+}(x)  \phi _{-}(x)\right)^+$, can
be obtained with the help of the complex conjugation of expression
(15). Eqs. (14), (15) can be used for the normalization of the
wave function.

\section{The soliton-like electric impulse}

Let us consider Eq. (6) for a particle in the field of the
electric impulse field. The non-stationary spatially homogeneous
electric field is defined by the 4-vector potential
\begin{equation}
A_\mu =\left( 0,0,A_3 (t),0\right),\hspace{0.3in}A_3 (t)=-a_0
\tanh k_0 t , \label{16}
\end{equation}
so that the electric field represents the soliton-like electric
impulse:
\begin{equation}
E_3 =\frac{a_0 k_0}{\cosh^2 k_0 t} , \label{17}
\end{equation}
$\textbf{E}=(0,0,E_3)$. The problem of the pair production of
particles with the spin $1/2$ by the field (17) was considered in
\cite{Narozhnyi}, \cite{Marinov}. Here we generalize this result
on the case of arbitrary spin particles. From Eq. (10) with the
help of Eqs. (16)-(17), we arrive at the equation
\[
\biggl[\frac{d^2}{dt^2}+p_0^2+2ep_3 a_0 \tanh k_0 t+e^2 a_0^2
\tanh^2 k_0 t
\]
\vspace{-8mm}
\begin{equation}  \label{18}
\end{equation}
\vspace{-8mm}
\[
+ \frac{ie\epsilon ga_0 k_0}{\cosh^2 k_0
t}S_3\biggr]\Psi_{\epsilon} (t)=0 ,
\]
where $p_0^2=m^2+\textbf{p}^2$ ,
\[
\psi_{\epsilon} (x)=\Psi_{\epsilon} (t)\exp (i\textbf{p x}) .
\]
The equations (18) with the representations $\epsilon=1$ and
$\epsilon=-1$ are connected by the complex conjugation.

We note that when the transverse momentum of particles approaches
to zero, $p_\bot\rightarrow 0$ ($p_\bot^2=p_1^2+p_2^2$), the spin
operator (7), for the field (16), becomes $\hat{S}_
{\textbf{p}}=J_3$. Then according to Eq. (9) the wave function
$\Psi_{\epsilon} (t)$ obeys the equation
\begin{equation}
S_3 \Psi_{\epsilon} (t)=s_z \Psi_{\epsilon} (t) . \label{19}
\end{equation}

Taking into consideration Eq. (19), one of solutions to Eq. (18)
is given by
\begin{equation}
{}_{-}\Psi^{(s_z)}_{\epsilon}(t)={}_{-}N(-z)^{i\mu}
\left(1-z\right)^{\lambda}
F(i\mu-i\nu+\lambda,i\mu+i\nu+\lambda;1+2i\mu;z)\Psi_0^{(s_z)} ,
\label{20}
\end{equation}
where $F(a,b;c;z)$ is the hypergeometric function \cite{Bateman},
$\Psi_0^{(s_z)}$ is the constant normalized eigenvector obeying
Eq. (19), $_{-}N$ is the normalization constant, and parameters
$\mu$, $\nu$, $\lambda$, $z$ are
\[
z=\exp (2k_0 t) ,~~~~2k_0\mu=\sqrt{p_\bot^2+m^2+(p_3-ea_0)^2}
\]
\begin{equation}
2k_0\nu=\sqrt{p_\bot^2+m^2+(p_3+ea_0)^2} , \label{21}
\end{equation}
\[
\lambda=\frac{1}{2}+\sqrt{\frac{1}{4}-\frac{e^2
a_0^2}{k_0^2}+i\epsilon s_z g\frac{ea_0}{k_0}} .
\]
Solution (20) has the following asymptotic limit at $t\rightarrow
\pm \infty $:
\[
\lim_{t\rightarrow +\infty}{}_{-}\Psi
^{(s_z)}_{\epsilon}(t)={}_{-}N\left[C^{(s_z)}_1 \exp (2i\nu k_0 t)
+ C^{(s_z)}_2 \exp(-2\nu k_0 t)\right]\Psi_0^{(s_z)} ,
\]
\vspace{-8mm}
\begin{equation}  \label{22}
\end{equation}
\vspace{-8mm}
\[
\lim_{t\rightarrow -\infty}{}_{-}\Psi
^{(s_z)}_{\epsilon}(t)={}_{-}N\exp (2i\mu k_0 t)\Psi_0^{(s_z)} ,
\]
where coefficients $C^{(s_z)}_1$ and $C^{(s_z)}_2$ are given by
\[
C^{(s_z)}_1=\frac{\Gamma
(2i\nu)\Gamma(1+2i\mu)}{\Gamma(i\mu+i\nu+\lambda)\Gamma(1+i\mu+i\nu-\lambda)}
,
\]
\vspace{-8mm}
\begin{equation}  \label{23}
\end{equation}
\vspace{-8mm}
\[
C^{(s_z)}_2=\frac{\Gamma
(-2i\nu)\Gamma(1+2i\mu))}{\Gamma(i\mu-i\nu+\lambda)\Gamma(1+i\mu-i\nu-\lambda)}
,
\]
and $\Gamma (x)$ is the Gamma-function. We imply here that the
transverse momentum of particles approaches to zero,
$p_\bot\rightarrow 0$. Equation (22) indicates that the solution
$_{-}\Psi ^{(s_z)}_{\epsilon}(t)$ corresponds to the negative
frequency at $t\rightarrow -\infty$ (that is why we use the
subscript $-$ in the wave function $_{-}\Psi
^{(s_z)}_{\epsilon}(t)$). It also follows from Eqs. (22) that
$2k_0\mu$ and $2k_0\nu$ are the kinetic energies of particles at
$t\rightarrow -\infty$ and $t\rightarrow +\infty$, respectively
\cite{Narozhnyi}, and the function
$_{-}\Psi^{(s_z)}_{\epsilon}(t)$ contains only negative frequency
at $t\rightarrow -\infty$. The solution $_{+}\Psi
^{(s_z)}_{\epsilon}(t)$ to Eq. (18) with the positive frequency at
$t\rightarrow -\infty$ may be obtained from Eq. (20) by the
substitution $\mu\rightarrow -\mu$:
\begin{equation}
{}_{+}\Psi ^{(s_z)}_{\epsilon}(t)={}_{+}N(-z)^{-i\mu}
\left(1-z\right)^{\lambda}
F(-i\mu-i\nu+\lambda,-i\mu+i\nu+\lambda;1-2i\mu;z)\Psi_0^{(s_z)} ,
\label{24}
\end{equation}
so that
\begin{equation}
\lim_{t\rightarrow -\infty}{}_{+}\Psi ^{(s_z)}_{\epsilon}
(t)={}_{+}N\exp (-2i\mu k_0 t)\Psi_0^{(s_z)} . \label{25}
\end{equation}

Eq. (18) is invariant, as well as equation for spin $1/2$
\cite{Narozhnyi}, under the replacement: $t\rightarrow -t$,
$e\rightarrow -e$, $s_z\rightarrow -s_z$ ($\mu \leftrightarrow
\nu$, $\lambda \rightarrow \lambda$). Making this substitution in
Eq. (20), we find the solution to Eq. (18) with the negative
frequency at $t\rightarrow \infty$:
\[
{}^{-}\Psi^{(s_z)}_{\epsilon}(t)={}^{-}N(-z)^{i\nu}
\left(1-z^{-1}\right)^{\lambda}
F(i\nu-i\mu+\lambda,i\nu+i\mu+\lambda;1+2i\nu;z^{-1})\Psi_0^{(s_z)}
,
\]
\vspace{-8mm}
\begin{equation}  \label{26}
\end{equation}
\vspace{-8mm}
\[
\lim_{t\rightarrow \infty}{}^{-}\Psi
^{(s_z)}_{\epsilon}(t)={}^{-}N\exp (2i\nu k_0 t)\Psi_0^{(s_z)} .
\]
In the same manner, using the replacement $\nu\rightarrow -\nu$,
we obtain from Eq. (26) the solution with the positive frequency
\[
{}^{+}\Psi ^{(s_z)}_{\epsilon}(t)={}^{+}N(-z)^{-i\nu}
\left(1-z^{-1}\right)^{\lambda}
F(-i\nu-i\mu+\lambda,-i\nu+i\mu+\lambda;1-2i\nu;z^{-1})\Psi_0^{(s_z)}
,
\]
\vspace{-8mm}
\begin{equation}  \label{27}
\end{equation}
\vspace{-8mm}
\[
\lim_{t\rightarrow \infty}{}^{+}\Psi
^{(s_z)}_{\epsilon}(t)={}^{+}N\exp (-2i\nu k_0 t)\Psi_0^{(s_z)} .
\]

With the help of Eqs. (22), (27) one can obtain the asymptotic (at
$t\rightarrow \infty$) relation
\begin{equation}
{}_{-}\Psi ^{(s_z)}_{\epsilon}(\infty)=a^{(s_z)}~{}^{-}\Psi
^{(s_z)}_{\epsilon}(\infty)+b^{(s_z)}~{}^{+}\Psi
^{(s_z)}_{\epsilon}(\infty) , \label{28}
\end{equation}
where
\begin{equation}
a^{(s_z)}=\frac{{}_{-}N}{{}^{-}N} C^{(s_z)}_1
,~~~~b^{(s_z)}=\frac{{}_{-}N}{{}^{+}N} C^{(s_z)}_2 . \label{29}
\end{equation}
It follows from Eqs. (2), (5), (12), (28) that the wave function
of the first order RWE (1) obeys at $t\rightarrow \infty$ the same
equation as Eq. (28):
\begin{equation}
{}_{-}\phi ^{(s_z)}_{\epsilon}(x)=a^{(s_z)}~{}^{-}\phi
^{(s_z)}_{\epsilon}(x)+b^{(s_z)}~{}^{+}\phi ^{(s_z)}_{\epsilon}(x)
,\label{30}
\end{equation}
where functions ${}_{-}\phi ^{(s_z)}_{\epsilon}(x)$ (with the
negative frequency at $t\rightarrow -\infty$), ${}^{-}\phi
^{(s_z)}_{\epsilon}(x)$ (with the negative frequency at
$t\rightarrow \infty$), ${}^{+}\phi ^{(s_z)}_{\epsilon}(x)$ (with
the positive frequency at $t\rightarrow \infty$) satisfy Eq. (9).

The pair production probability of particles by electromagnetic
fields can be obtained through the asymptotic of the exact
solutions of wave equations \cite{Nikishov}. Relation (30) and
complex conjugated equation generate the Bogolyubov
transformations of creation and annihilation operators.
Coefficients $a^{(s_z)}$ and $b^{(s_z)}$, Eq. (29), contain
information about pair creation and obey the relations
\[
|a^{(s_z)}|^2 -|b^{(s_z)}|^2 =1 ~~~~ \mbox{for~bosons} ,
\]
\vspace{-8mm}
\begin{equation}  \label{31}
\end{equation}
\vspace{-8mm}
\[
|a^{(s_z)}|^2 +|b^{(s_z)}|^2 =1 ~~~~ \mbox{for~fermions} .
\]
According to the approach \cite{Nikishov}, the density of pairs of
arbitrary spin particles created during all time is given by
\begin{equation}
n^{(s)}=\sum_{s_z}\int \mid b^{(s_z)}\mid^2\frac{d^3p}{(2\pi)^3} .
\label{32}
\end{equation}

The formula (32) generalizes the result \cite{Nikishov} on the
case of arbitrary spin particles. The difference of Eqs. (23),
(32) from expressions for spin zero and one half \cite{Nikishov}
is in the parameter $\lambda$, Eq. (21) and values $a^{(s_z)}$ and
$b^{(s_z)}$.

\subsection{Spin-$0$, and $1/2$ particles}

Let us consider a particular case of spinless particles, $s=0$.
One may get the Klein -- Gordon -- Fock equation by setting
$s_z=0$ in Eq. (18). According to Eq. (15), normalization
constants in Eq. (29) depend on the spin of particles. For
spin-zero, one has to use the normalization constants
\cite{Nikishov}:
\begin{equation}
{}_{+}N^{(0)}={}_{-}N^{(0)}=\left(4k_0\mu\right)^{-1/2} ,
~~~~{}^{+}N^{(0)}={}^{-}N^{(0)}=\left(4k_0\nu\right)^{-1/2} .
 \label{33}
\end{equation}
Putting $s_z=0$ in Eqs. (21), (23), (29), and using the formulas
\cite{Bateman}:
\begin{equation}
\mid \Gamma (1+iy)\mid^2=\frac{\pi y}{\sinh\pi y} ,~~~~ \Gamma (z)
\Gamma (1-z)=\frac{\pi}{\sin\pi z} , \label{34}
\end{equation}
we obtain
\[
\lambda =\frac{1}{2}+ \sqrt{\frac{1}{4}-\frac{e^2 a_0^2}{k_0^2}} ,
\]
\begin{equation}
|a^{(0)}|^2 = \frac{\sin \pi(i\mu +i\nu +\lambda)\sin \pi(-i\mu
-i\nu +\lambda)}{\sinh2\pi\mu\sinh2\pi\nu} , \label{35}
\end{equation}
\[
 |b^{(0)}|^2
=\frac{\sin \pi(i\mu -i\nu +\lambda)\sin \pi(-i\mu +i\nu
+\lambda)}{\sinh2\pi\mu\sinh2\pi\nu} .
\]
The Bogolubov coefficients $a^{(0)}$, $b^{(0)}$, Eq. (35), obey
the relation (31) for bosons. With the help of Eqs. (32), (35) one
can get the known result \cite{Nikishov} (see also \cite{Grib}):
\begin{equation}
n^{(0)}=\int \frac{\cos^2 \pi\vartheta ' +\sinh^2 \pi
(\mu-\nu)}{\sinh 2\pi\mu \sinh 2\pi\nu} \frac{d^3p}{(2\pi)^3}
,~~~~\vartheta'=\sqrt{\frac{1}{4}-\frac{e^2 a_0^2}{k_0^2}} .
\label{36}
\end{equation}

Consider the case of spin-1/2 particles. Putting $s_z= 1/2$,
$g=2$, we find from Eqs. (21), (23), (34)
\[
\lambda =1+ i\vartheta ,~~~~\vartheta=\frac{e a_0}{k_0} ,
\]
\begin{equation}
|C^{(1/2)}_1|^2 = \frac{\mu(\mu+\nu-\vartheta)\sinh \pi(\mu +\nu
+\vartheta)\sinh \pi(\mu +\nu
-\vartheta)}{\nu(\mu+\nu+\vartheta)\sinh2\pi\mu\sinh2\pi\nu} ,
\label{37}
\end{equation}
\[
|C^{(1/2)}_2|^2 =\frac{\mu(\mu-\nu-\vartheta)\sinh \pi(\mu -\nu
+\vartheta)\sinh \pi(\mu -\nu
-\vartheta)}{\nu(\mu-\nu+\vartheta)\sinh2\pi\mu\sinh2\pi\nu} .
\]
Equations (37) correspond to the spin projection $s_z=1/2$ and the
representation with the parameter $\epsilon=1$. For the spin
projection $s_z=-1/2$ one should make the replacement
$\vartheta\rightarrow -\vartheta$ in Eqs. (37). Using the
normalization constants \cite{Nikishov}
\[
{}_{+}N^{(1/2)}=\left[8k_0\mu\left(2k_0\mu
-p_3+ea_0\right)\right]^{-1/2} ,
\]
\[
{}_{-}N^{(1/2)}=\left[8k_0\mu\left(2k_0\mu
+p_3-ea_0\right)\right]^{-1/2} ,
\]
\vspace{-8mm}
\begin{equation}  \label{38}
\end{equation}
\vspace{-8mm}
\[
{}^{+}N^{(1/2)}=\left[8k_0\nu\left(2k_0\nu
-p_3-ea_0\right)\right]^{-1/2} ,
\]
\[
{}^{-}N^{(1/2)}=\left[8k_0\nu\left(2k_0\nu
+p_3+ea_0\right)\right]^{-1/2} ,
\]
and the relations
\begin{equation}
\frac{2k_0\nu -p_3-ea_0}{2k_0\mu
+p_3-ea_0}=\frac{\mu-\nu+\vartheta}{\nu-\mu +\vartheta} ,~~~~
\frac{2k_0\nu +p_3+ea_0}{2k_0\mu
+p_3-ea_0}=\frac{\mu+\nu+\vartheta}{\nu+\mu -\vartheta}
,\label{39}
\end{equation}
one may find the coefficients
\[
a^{(1/2)}=\frac{\sinh \pi(\mu +\nu +\vartheta)\sinh \pi(\mu +\nu
-\vartheta)}{\sinh2\pi\mu\sinh2\pi\nu} ,
\]
\vspace{-8mm}
\begin{equation}  \label{40}
\end{equation}
\vspace{-8mm}
\[
b^{(1/2)}=\frac{\sinh \pi(\nu -\mu -\vartheta)\sinh \pi(\mu -\nu
-\vartheta)}{\sinh2\pi\mu\sinh2\pi\nu} .
\]
Coefficients $a^{(1/2)}$, $b^{(1/2)}$, Eq. (40), satisfy Eq. (31)
for fermions. We recover from Eqs. (29), (32), (40) the density of
electron-positron pairs created by the electric field (17)
\cite{Nikishov}:
\begin{equation}
n^{(1/2)}=2\int \frac{\sinh \pi(\nu-\mu -\vartheta)\sinh
\pi(\mu-\nu -\vartheta)}{\sinh 2\pi\mu \sinh 2\pi\nu}
\frac{d^3p}{(2\pi)^3} . \label{41}
\end{equation}
The coefficient 2 in Eq. (41) is due to two spin pojections
$s_z=\pm 1/2$.

\section{Constant and uniform electric field}

We can get the approximation of a constant field by considering
the limiting case $ea_0/m\rightarrow \infty$ \cite{Nikishov},
\cite{Grib}. Setting $a_0=E/k_0$, one arrives from Eq. (16) at
$k_0\rightarrow 0$:
\begin{equation}
A_3 =-\frac{E}{k_0}\tanh k_0 t\rightarrow -Et .
 \label{42}
\end{equation}
The vector-potential (42) describes the uniform and constant
electric field. From Eq. (21) we find the approximate relations
\[
\mu\simeq \nu\simeq \frac{eE}{2k_0^2}+\frac{p_\bot^2 +m^2}{4eE} ,
\]
\vspace{-8mm}
\begin{equation}  \label{43}
\end{equation}
\vspace{-8mm}
\[
\lambda\simeq\frac{1}{2}\left(1+s_z g\right)+i\frac{eE}{k_0^2} .
\]
With the help of asymptotic formula \cite{Bateman}
\begin{equation}
|\Gamma
(x+iy)|=\sqrt{2\pi}|y|^{x-1/2}\exp\left(-\frac{\pi|y|}{2}\right)~~~~\mbox{at}~|y|\rightarrow
\infty ,
 \label{44}
\end{equation}
we find the approximate relations
\[
|\Gamma \left(i\mu+i\nu
+\lambda\right)|^2=2\pi\left(\frac{eE}{k_0^2}\right)^{s_z
g}\exp\left(-\frac{\pi eE}{k_0^2}\right) ,
\]
\vspace{-7mm}
\begin{equation}  \label{45}
\end{equation}
\vspace{-7mm}
\[
|\Gamma \left(1+i\mu+i\nu
-\lambda\right)|^2=2\pi\left(\frac{eE}{k_0^2}\right)^{-s_z
g}\exp\left(-\frac{\pi eE}{k_0^2}\right) .
\]
Using Eqs. (34), (45) it is not difficult to obtain from Eq. (23)
the expression
\begin{equation}
|C_2^{(s_z)}|^2= \exp\left(-\frac{\pi\left(p^2_\bot
+m^2\right)}{eE}\right) . \label{46}
\end{equation}
For the approximation considered the Bogolyubov coefficient (29)
$b^{(s_z)}=C_2^{(s_z)}$. As a result, from Eq. (32) one obtains
the density of arbitrary spin particles pairs created during all
time
\begin{equation}
n^{(s)}=\sum_{s_z}\int \exp\left[-\frac{\pi\left(p^2_\bot
+m^2\right)}{eE}\right]\frac{d^3p}{(2\pi)^3} .
 \label{47}
\end{equation}
 For the constant and uniform electric field, according to the approach
 \cite{Nikishov}, one should make the replacement
\begin{equation}
\int dp_3 \rightarrow eET ,
 \label{48}
\end{equation}
where $T$ is infinite time of observation. Inserting the
replacement (48) into Eq. (47) and calculating integral, we obtain
the intensity of pair production of arbitrary spin particles by
the constant and uniform electric field
\begin{equation}
I(E)=
\frac{n^{(s)}}{T}=\left(2s+1\right)\frac{e^2E^2}{8\pi^3}\exp\left(-\frac{\pi
m^2}{eE}\right) .
 \label{49}
\end{equation}
Expression (48) is in agreement with the differential probability
calculated on the basis of exact solutions of equations for
arbitrary spin particles in the external constant and uniform
electromagnetic fields \cite{Kruglov2001}. The function (49)
possesses the nonanalytical dependance on the electric field,
which indicates that the intensity (49) can not be received by
perturbative theory.

The intensity of pair production of arbitrary spin particles, Eq.
(49), is ($ 2s+1)$ times that for the scalar particle intensity
due to the ($2s+1)$ physical degrees of freedom of the arbitrary
spin field. The probability of pair production decreases rapidly
with the mass of particles because of the exponential factor.

\section{Conclusion}

The number of pairs of arbitrary spin particles produced by a
uniform soliton-like electric field was obtained on the basis of
exact solutions. The intensity of pair production was expressed
through the Bogolyubov coefficient $C_2$ which with the
coefficient $C_1$ relate the asymptotics of exact solutions at
$t\rightarrow \pm \infty$. At a particular case ($k_0\rightarrow
0$), when the field is converted into constant electric field, we
come to our earlier result. This case corresponds to the adiabatic
regime \cite{Popov} which can be now achieved experimentally. In
another case, with the small pulse duration, the probability can
increase considerably.

For spin $0$ and $1/2$, the probabilities of pair production
reduce to expressions found in \cite{Nikishov}. Although the light
electron-positron pairs are created by the smallest value of the
electric field, the general case of arbitrary spin particles is of
definite theoretical interest.

Pair production of particles are described by nonperturbative
effects of electrodynamics and it is impossible to obtain them by
summing of definite terms of perturbation theory. Therefore, the
observation of the pair production by the external electromagnetic
field experimentally is of great scientific interest.


\begin{thebibliography}{99}

\bibitem{Schwinger}  J. Schwinger. Phys. Rev. \textbf{82} (1951), 664.
\bibitem{Marinov72}  M. S. Marinov and V. S. Popov, Yad. Fiz. \textbf{15} (1972),
1271 [Sov. J. Nucl. Phys. \textbf{15} (1972), 702].
\bibitem{Kruglov2001} S. I. Kruglov, Ann. Phys. (N.Y.) \textbf{293} (2001), 228;
arXiv:hep-th/0110061; Eur. Phys. J. {\bf C22} (2001), 89;
arXiv:hep-ph/011010.
\bibitem{SLAC} CERN Courier \textbf{40} (6) (2000), 26; \textbf{41} (5) (2001),
20.
\bibitem{Ringwald} A. Ringwald, Preprint DESY 01-213 (2001);
Phys. Lett. \textbf{B 510 }(2001), 107.
\bibitem{Alkofer} R. Alkofer et.al., Phys. Rev. Lett. \textbf{87} (2001), 193902.
\bibitem{Popov} V. S. Popov, Sov. Phys.-JETP \textbf{94} (2002), 1057 [Zh.
Eksp. Teor. Fiz. \textbf{121} (2002), 1235]
\bibitem{Hurley}  W. J. Hurley. Phys. Rev. D\textbf{4} (1971), 3605; D\textbf{10}
(1974), 1185; Phys. Rev. Lett. \textbf{29} (1972), 1475.
\bibitem{Kruglov}  S. I. Kruglov, Int. J. Theor. Phys. \textbf{40} (2001), 515;
arXiv:hep-ph/9908410.
\bibitem{Feynman} R. P. Feynman and M. Gell-Mann, Phys. Rev. \textbf{109} (1958),
193;
\bibitem{Broun} L. M. Broun, Phys. Rev. \textbf{111} (1958), 957.
\bibitem{Narozhnyi}  N. B. Narozhnyi, A. I. Nikishov, Sov. J. Nucl. Phys. \textbf{11}
(1970), 596.
\bibitem{Marinov}  M. S. Marinov and V. S. Popov, Fortschr. Phys. \textbf{25} (1972),
401
\bibitem{Bateman}  H. Bateman and A. Erdelyi, Higher Transcendental Functions, New
York, McGraw-Hill, 1953.
\bibitem{Nikishov}  A. I. Nikishov, Sov. Phys.-JETP \textbf{30} (1970), 660 [Zh.
Eksp. Teor. Fiz. \textbf{57} (1969), 1210]; Nucl. Phys.
B\textbf{21 }(1970), 346; Trudy FIAN, V.111, p.152, Nauka, Moscow,
1979 [J. Sov. Laser Res. \textbf{6} (1985), 619].
\bibitem{Grib}  A. A. Grib, S. G. Mamayev and V. M. Mostepanenko, Quantum
Effects in Intensive External Fields (Moscow: Atomizdat, 1980).

\end{thebibliography}
\end{document}